# The Development of the College Park Tornado of 24 September 2001

By

KENNETH L. PRYOR

ABSTRACT

The 24 September 2001 College Park, Maryland, tornado was remarkable because of its long-track that passed within a close range of two Doppler radars. This tornado featured many similarities to previous significant tornado events that resulted in widespread damage in urban areas, such as the Oklahoma City tornado of 3 May 1999. The College Park tornado was the third in a series of three tornadoes associated with a supercell storm that developed over central Virginia. It was initiated 3 - 4 km southwest of College Park and dissipated over eastern Howard County, near Columbia. The supercell that produced the tornado tracked approximately 120 km from Stafford County, Virginia to eastern Howard County in about 126 minutes.

This paper presents a synoptic and mesoscale overview of favorable conditions and forcing mechanisms that resulted in the severe convective outbreak associated with the College Park tornado. Convective morphology will be examined in terms of Doppler radar and satellite imagery. MM5 output data and GOES imagery on 24 September revealed many critical elements of the tornadic event, including a negative-tilted upper-level short-wave trough over the Ohio Valley, a jet stream with strong vertical shear, and a warm, moist tongue of the air associated with strong southerly flow over central Maryland and Virginia during the mid-afternoon. Operational WSR-88D and Terminal Doppler Weather Radar (TDWR) data showed a high reflectivity knob within the hook echo, the evolution of the parent storm from a supercell structure to a bow echo, and a tornadic cyclone signature. Many of the features identified in the MM5 data were verifiable by observations. In addition, a major non-observable feature identified in the MM5 data was a solitary wave associated with a density current that was a likely trigger for deep convection. This study concludes with a discussion of the effectiveness of using MM5 guidance in conjunction with satellite and radar imagery in the operational environment of forecasting severe convection.

## Acknowledgements

The author wishes to thank Mr. Steven Zubrick of the National Weather Service Forecast Office, Sterling, Virginia) for providing a comprehensive radar data set for analysis in this research effort. Dr. Weizhong Zheng (University of Maryland) provided technical support and guidance in the generation of model simulations featured in this paper.

# Chapter 1. Introduction

**1. The College Park tornado**

A remarkable F-3 tornado struck College Park, Maryland, hereafter referred to as the College Park tornado, on 24 September 2001 with its track passing within a close range of two Doppler radars. The College Park tornado was the third in a series of three tornadoes produced by a supercell storm that developed from the splitting of a pre-existing supercell at its right flank around 1944 UTC over Stafford County, Virginia. Its left-flank storm was weaker than the right-flank supercell and eventually dissipated by 2029 UTC. However, the right-flank supercell was maintained as a quasi-steady storm in which three tornadoes were spawned. The first tornado (F0 intensity) had an 18-km path through Stafford and Prince William Counties between 2010 and 2032 UTC and touched down on Quantico Marine Base (see Fig. 1). The second tornado (F1 intensity) had a 25-km path from Fort Belvior, Virginia to the U.S. Capitol in the District of Columbia, between 2044 and 2112 UTC. The College Park tornado, the third and most intense one (F3 intensity), developed 3-4 km southwest of College Park at approximately 2116 UTC and eventually dissipated over eastern Howard County by 2150 UTC. The supercell that spawned the College Park tornado tracked approximately 120 km from Stafford County, Virginia to eastern Howard County, Maryland in about 126 minutes. Throughout its 120-km track, radar reflectivity shows a distinct hook echo associated with the supercell.

The College Park tornado produced heavy damage from just west of the campus of University of Maryland to downtown Laurel. On and near the campus, the tornado caused approximately $15 million in damage, including 10 destroyed trailers (used as temporary facility for the Maryland Fire and Rescue Institute), several heavily damaged

buildings, many tossed and destroyed vehicles and two fatalities. The tornado then struck the U.S. Department of Agriculture Research Center, causing an estimated $41 million in damage to buildings and research documents. It continued north through Beltsville, damaging homes, businesses, and schools. In downtown Laurel, the tornado tore the roof off a wing of the Laurel High School and then destroyed a single-level house in the neighborhood behind the school. Moving from Prince Georges County into Howard County, the tornado caused damage to 43 homes before finally dissipating in Columbia. Overall, $16 million in damage resulted in Prince Georges County that included 861 residential homes, 560 vehicles, and at least 23 commercial businesses. Approximately an additional $1 million in damage was caused in Howard County. Total damage in Maryland was estimated to be over $73 million.

Tornado climatology as compiled by the National Climatic Data Center (NCDC) indicates that, on average, one strong to violent (F2 to F5 intensity damage) tornado occurs in the state of Maryland each year. This climatology is based on data collected between the years 1950 and 1995. The occurrence of the College Park tornado and the 28 April 2002 La Plata, Maryland tornado of F4 intensity within a one-year period possibly indicates a significant increasing trend in tornado occurrence in this region. Based on this apparent increasing trend, the U.S. Congress considered tasking the National Weather Service to initiate a study on the occurrence of the two tornadoes. The purpose of this study, as stated in the Maryland Tornado Study Team Charge drafted by Dr. P. Hirschberg (NWS), was to "determine whether the recent occurrence of tornadoes in the state of Maryland is indicative of an increasing trend in these events and if so, what this trend is attributable to". Also addressed in this study was the question of whether this increasing trend is resulting from natural and/or anthropogenic climate change.

In addition to the intensity of the College Park tornado and the damage it produced, another curious aspect of this event was the real-time, operational forecast of the mesocyclone associated with the tornado. That is, the real-time forecast using the Pennylvania State University-National Center for Atmospheric Research fifth generation mesoscale, nonhydrostatic, nested-grid model, referred to as the MM5, identified a vortex associated with the mesocyclone. The predicted timing and location of this vortex, especially apparent in the MM5-predicted radar reflectivity, was within 2400 s (40 minutes) and 20 km of the observed, respectively. The model predictions were made on three nested domains with a grid size of 36 km in the outer domain, a grid size of 12 km in the next inner domain and a grid size of 4 km in the inner-most domain. The inner domain (12 km grid size) and the inner-most domain (4 km grid size) use progressively finer resolution terrain data. To help gain insight into the small-scale features of the storm, the MM5 was rerun for the present study with all the model settings identical to the then real-time forecast except that the finest grid of 1.33 km was nested inside the 4-km resolution domain. The model utilizes a terrain-following (sigma) coordinate with 24 layers and a radiative upper-boundary condition at 50 hPa.

## 2. Previous research

There has been considerable progress in the observational studies and understanding of various facets of tornadoes. An example of worth mentioning is the development of a large tornado that occurred around Oklahoma City on 3 May 1999. This case has generated a number of publications, and may serve as a model for a strong tornado that had a major impact on a metropolitan area. For example, Burgess et al.

(2002) identified the importance of analyzing data from multiple radars at close range to such a large tornado, which provided an opportunity to investigate the structure and evolution of the tornado and the rotational flows surrounding it as well as the relationship of the radar signatures to tornado intensity.

Bikos et al. (2002) identified important features causing convective morphology from GOES imagery, including a deepening upper-level trough over the western United States and a rapidly propagating jet streak. Differing cloud fields were utilized to delineate air masses and boundaries. Several key findings, identified in visible imagery, were related to the outbreak of severe convection: (a) developing and dissipating cumulus convection in the dry air west of a dryline over western Texas as the leading edge of jet stream cirrus moved into the area; (b) towering cumulus convection developing in southwestern Oklahoma that would eventually evolve into tornadic supercell activity in the Oklahoma City area; (c) a pre-existing mesoscale boundary separating a region of cumulus cloud lines in an unstable air mass from a stable region characterized by wave cloudiness and its interaction with the evolving supercell storm; (d) the development of the first storm over southwestern Oklahoma, its rapid intensification and evolution into a tornadic supercell, and its movement along the mesoscale boundary toward Oklahoma City, as displayed by animated imagery.

Foster et al. (2000) analyzed the evolution of surface meteorological fields and found four distinct air masses across the southern Plains that contributed to convective initiation and organization. The four air masses included: (a) cool, moist air mass produced by heavy rainfall that occurred on the previous day in north Texas that was advected by southerly flow into central Oklahoma during the day of 3 May; (b) a warm, moist, air mass west of a previously existing dry line segment over western Oklahoma and western Texas that had resulted from diabatic heating due to solar radiation and the subsequent evaporation of antecedent rainfall; (c) a dry air mass over extreme western Texas and the Oklahoma Panhandle characterized by deep mixing that resulted in the development of a new dry line segment during the afternoon of 3 May; and (d) a dry air mass originating over southwestern Texas characterized by mixing of deep moisture and momentum that surged northeastward during the afternoon of 3 May, resulting in the development of a bulging dryline segment over west Texas. The cool, moist air mass over central Oklahoma and the warmer, more unstable air mass over western Oklahoma resulted in the establishment of the mesoscale boundary, previously discussed in Bikos et al. (2002) that served to enhance convergence and the intensity and longevity of the tornadic supercell storm that struck the Oklahoma City area. The bulging dryline segment further to the southwest over western Texas initiated additional deep, moist convection. Addition of mesonet data into the analysis revealed more detail: a mesoscale low near the dryline bulge over western Texas as well as low pressure trough extending east of the low. Most importantly, the surface moisture flux divergence field revealed a large area of moisture convergence over southwest Oklahoma where convective initiation occurred. The convergence along the dry line segments and the mesoscale boundary over central Oklahoma in an unstable air mass characterized by strong vertical wind shear enhanced the favorability for intense, rotating updrafts and the development of tornadic supercellular convection. These surface features also interacted with upper-tropospheric features such as a jet streak and vorticity maximum to enhance convective updrafts.

    Roebber et al. (2002) utilized the MM5 in forecast mode to explore the sensitivity

of the 3 May 1999 tornado outbreak to several features, and found that: (a) convective initiation in the weakly forced environment was achieved through modification of the existing cap by both surface heating and synoptic scale ascent associated with the PV anomaly; (b) supercellular organization was supported regardless of the strength of the southern PV anomaly; and (c) the cirrus shield was important in limiting development of convection and reducing competition between storms.

Research of other tornado outbreaks has also identified specific features relevant to the study of tornadic supercell storm structures. For example, research from the tornado events of the Union City, Oklahoma tornado of 24 May 1973, the Stillwater, Oklahoma tornado of 13 June 1975, and the Binger, Oklahoma tornado of 22 May 1981, found characteristics similar to the College Park tornado. The Union City and Binger, Oklahoma tornadoes were both strong tornadoes (at least F4 intensity damage) that lofted large amounts of debris, and thus made possible the inference of tornado location as well as the flow field surrounding the tornado (Brown et al. 1978, Lemon et al. 1982). Doppler radar measurements in the Union City tornado resulted in the discovery of the tornadic vortex signature (TVS) in the field of mean Doppler velocity data because of its anomalous character relative to the surrounding mean velocity field. The TVS can be identified as a couplet of extreme mean Doppler velocity values that occur approximately one-half beamwidth on either side of the tornado when the sampling volume is centered at the same range as the tornado (Brown et al. 1978). It was also found in the study of the Union City tornado that the TVS signature could not be resolved unless the peak-to-peak Doppler velocity shear is appreciably greater than the background cyclonic shear produced by the parent mesocyclone. Similar to the supercell that produced the College

Park tornado, the storm that produced the Union City tornado was an isolated storm that developed ahead of a convective line. In the Union City case, the TVS coincided very closely with the surface damage path. In the Stillwater tornado case, the TVS was detectable at the lowest levels while the tornado was inflicting damage on the ground. The intense shear of the TVS was at the mesocyclone center, coinciding with the reported tornado location. The anomalous character of the Stillwater TVS, like the Union City TVS, distinguished it in the velocity field. In addition to the TVS identified in earlier tornado studies, the study of the Binger, Oklahoma tornado revealed a distinct reflectivity maximum in excess of 50 dBZ that coincided with the TVS near the surface. The low-level reflectivity maximum associated with the TVS in the Binger case may be related to the debris cloud composed of large missiles (Lemon et al. 1982).

**3. Objectives of this study**

The College Park tornado featured several similarities to previous significant tornado events causing widespread damage in urban areas. The objectives of the present scholarly paper are to

a) document favorable conditions and forcing mechanisms at both the synoptic and mesoscale that resulted in the severe convection outbreak and tornadic genesis using GOES imagery, upper-air observations from Dulles Airport, Virginia (KIAD) and MM5 output.

b) examine the morphology of the mesocyclone associated with the College Park tornado using the Doppler radar data obtained at Sterling , Virginia, and Baltimore Airport, Maryland.

c) validate the MM5 forecast data, initialized at 1200 UTC 24 September 2001,

against the radar and satellite imagery in order to assess its performance during the severe convection outbreak, and

d) diagnose some peculiar features that were not apparent in the observations (e.g. radar and satellite imagery), such as a solitary wave associated with a rapidly propagating density current.

The next chapter provides a larger-scale overview of the upper-level and surface features that are favorable for severe convection and tornadic activity utilizing satellite imagery and MM5 output data. Chapter 3 shows the structure and evolution of the tornado and its parent supercell, as well as convective morphology using high-resolution Doppler radar and satellite imagery. Chapter 4 presents the prediction of the tornado-related events using MM5. A summary and concluding remarks are given in the final chapter.

# Chapter 2. Larger-Scale Overview

1. **Synoptic overview**

Figure 3 shows the MM5's 12-h forecast overlying GOES water vapor imagery near 0000 UTC 25 September 2001. A negative-tilted upper-level short wave trough extended from the eastern Great Lakes region to the Delaware Bay. Note the presence of a wide belt of moisture ahead of the trough that was associated with a broad quasi-geostrophic ascent due to the presence of positive differential vorticity advection and warm advection. Moreover, widespread deep convection extended from New York to the eastern North Carolina.

Regional, high-resolution visible imagery was available every 30 minutes from 1715 to 2145 UTC, which allowed for tracking the evolution of the tornado-producing supercell. Figure 2a, visible imagery at 1915 UTC, displays transverse bands in cirrus clouds over the western Virginia and West Virginia, which are typical for jet streams (Conway 1997). The transverse bands occur at a direction perpendicular to the jet stream and, in this case, indicate strong southwesterly flow aloft. The MM5 forecast data indicate that a wind speed maximum was most apparent over the region at 500 mb (not shown). Note that College Park was in a favorable region of upward motion in addition to the above-mentioned quasi-geostrophic forcing.

Imagery at 1915 UTC also shows breaks in the cirrus deck associated with widespread cumulus congestus, and cumulus lines over the east-central Maryland and the eastern shore, respectively (Fig. 2a), revealing the presence of conditional instability beneath the jet core prior to the tornadic event. In accordance with the satellite imagery, an analysis of the MM5 output shows similar conditions (see Figs. 1, 4, and 5). Namely, the large scale and regional surface analyses (Figs. 1 and 4), and the 850-mb $\theta_e$ (Fig. 5) show the presence of a warm, moist and unstable tongue of air associated with strong southerly flow over central Maryland and Virginia during the mid-afternoon ahead of an eastward progressing cold front.

As is typical with northeastward-moving supercells, the low-level air approaches from the southeast and rises through the updraft. Thus, the updraft of the supercell storm that would eventually produce the College Park tornado was predominantly "fed" by a warm, moist and positively buoyant air mass from the right front quadrant of the storm (Rotunno 1986). The warm air and moist advection in the low levels result in the large positive buoyancy present over central and eastern Maryland. Note that the low-level southeasterly flow overlain by strong southwesterly flow aloft indicated the presence of strong directional wind shear that was to be a major contributor to the development of storm rotation and the favorability for splitting supercell storms.

A further analysis of the MM5 output data indicates conditions favorable for large-scale ascent over the eastern one-third of the United States during the afternoon of September 24. For example, the presence of a deep 500-mb trough (not shown) over the Ohio Valley implied positive differential vorticity advection (PDVA) over the eastern United States. The surface analysis at 2200 UTC indicates that low-level warm air advection (WAA) was occurring over central Maryland ahead of a frontal surface trough (see Fig. 4). The combined effects of PDVA and WAA resulted in large-scale ascent over central Maryland during the development of the College Park tornado as inferred from the omega equation. In the study of the synoptic regulation of the 3 May 1999 tornado outbreak, Roebber et al. (2002) identified that PDVA was important in the generation of midlevel synoptic-scale ascent that promoted deep convection. In addition, MM5-derived cross sectional imagery at 2200 UTC indicates that strong surface convergence was occurring over central Maryland prior to the arrival of the supercell (Fig. 5a). Foster et al. (2000), in their study of the evolution of surface features prior to the 3 May 1999 tornado outbreak, identified a large area of surface convergence over southwestern and central Oklahoma prior to convective initiation.

## 2. Mesoscale overview

Modified radiosonde observation (RAOB) at 2100 UTC from nearby Dulles Airport, Virginia (KIAD), shows a significant amount of buoyant energy, a moist layer from the surface to 700 mb and a well-defined dry layer aloft, that was favorable for severe convection and tornadic supercell development (Fig. 6). McGinley (1986) and Weisman and Klemp (1986) have identified that a dry air layer in the midlevels (650-500 mb) enhances storm severity by maximizing the vertical lapse rate of equivalent potential temperature ($\theta_e$), thereby increasing convective instability and parcel energy. Another means by which a mid level dry layer can enhance storm severity through the process of entrainment of dry air into the moist downdraft. This process results in evaporational cooling that increases the strength of the downdraft and storm outflow, thereby enhancing convergence along the gust front and subsequent updraft redevelopment (Weisman and Klemp 1986). In addition, the strong vertical wind shear was evident from the surface to approximately 700 mb. A comparison of the RAOBs between 2100 and 1200 UTC shows the backing of low-level winds as well as highlights the increasing potential for strong convection (Fig. 6). This resulting strong directional shear was a major contributor to the development of storm rotation and the favorability for splitting supercell storms.

This rotation could induce a vertical pressure gradient that might enhance and maintain the storm updraft (Weisman and Klemp 1986). In addition, the 2100 UTC KIAD RAOB displayed a considerable amount of buoyant energy available to "fuel" deep convection as indicated by the Convective Available Potential Energy (CAPE/B+) value of 1606 J/Kg. An analysis of the mesoscale model ensemble forecasts of the May 3, 1999 tornado outbreak (Weiss and Stensrud 2000) revealed the presence of an axis of strong CAPE over the central and southern Plains prior to the development of deep convection. The Lifted Index (LI) calculated from this sounding was -5, underscoring the strong convective instability of the ambient atmosphere. The Bulk Richardson Number (BRN) of 32 and the Severe Weather Threat Index (SWEAT) Index value of 402 were indicative of the presence of strong low-level directional wind shear and the potential for the development of tornadic supercell thunderstorms. The Bulk Richardson Number expresses a relationship between storm type, wind shear, and buoyancy and is represented by the formula

$$R = B/0.5U^2, \qquad (1)$$

where B is the buoyant energy in the storm's environment and U is a measure of the vertical wind shear (Weisman and Klemp 1982, Weisman and Klemp 1984). R values between 10 and 40 signify the potential for supercell development. The computed Total Totals (TT) Index of 51 indicated the potential for the development of severe thunderstorms and isolated tornadoes.

The MM5 output indicates the presence of a persistent $\theta_e$ ridge extending from central Maryland southward into Virginia (Fig. 5b). Chaston (1995) states that such a $\theta_e$ ridge, acting upon by a lifting mechanism (i.e., WAA, frontal lifting), can serve as an axis of available potential energy that is convertible into kinetic energy of the subsequent convection. In this case, the $\theta_e$ ridge provided a source of warm, moist and positively buoyant air to feed the supercell as it tracked northeastward into Maryland, producing strong convective updrafts. In addition, the cross section at 2200 UTC indicates a decrease of $\theta_e$ in the lower troposphere with a minimum value near 800 mb (Fig. 5a), a potentially unstable condition for deep convection over central Maryland prior to the arrival of the tornadic supercell. It follows that lifting of a potentially unstable layer can result in severe convection. The cross section also reveals that upward motion was increasing in strength and depth along and ahead of a frontal boundary as it tracked eastward across Virginia and central Maryland.

The model-generated soundings between 2000 and 2100 UTC are given in Fig. 7, which reveals gradual increases in CAPE and in vertical wind shear. As previously discussed, the large amount of buoyant energy resulted in the enhancement of the supercell updraft as it tracked into central Maryland. Increased updraft strength resulted in the upward tilting of horizontal vorticity associated with the large vertical wind shear, leading to the intensification of the vertical component of relative vorticity and cyclonically rotating updrafts (Rotunno 1986).

## Chapter 3. Radar and Satellite Morphology

Scans of the College Park tornado were available from the Sterling, Virginia WSR-88D (KLWX) radar and the Baltimore-Washington International Airport (KBWI) Terminal Doppler Weather Radar (TDWR). During its lifetime, the tornado was 39-59 km from KLWX and approximately 37-56 km from the BWI TDWR. The centerline heights of the lowest elevation angle beam were approximately 400-700 m AGL for the KLWX radar which, operated in volume coverage pattern (VCP) 11 (scans at elevation angles up to 19.5° with updates every 5 min), collected reflectivity and velocity data continuously during the tornado. The BWI TDWR, with an elevation scan of 0.3°, also collected data continuously during the event.

TDWR provides a higher resolution display compared to the WSR-88D by virtue of its radar characteristics (see Table 1). Since azimuth resolution is proportional to the antenna beam width of the radar system while range resolution is proportional to the pulse length, the narrower beamwidth and shorter pulse length of the TDWR compared to the WSR-88D result in higher azimuth and range resolution, respectively. Thus, TDWR can provide a more precise display of radar reflectivity signatures.

Figure 8 displays a well-defined hook echo within the supercell as it tracked from Washington, DC to eastern Howard County. The hook echo has been identified as an indicator of the existence of a mesocylone, defined as a 3-dimensional vortex in a convective storm that rotates cyclonically and is closely correlated with severe convective activity (Rinehart 1997). As the supercell tracked northeastward, the hook echo developed a prominent "knob" (Burgess et al. 2002) defining the location of the tornado (see Figs. 8c and 8d). Higher reflectivities in the knob (i.e., >55 dBZ) began at approximately 2121 UTC near College Park and continued beyond 2146 UTC, when the tornado was located near Columbia, Maryland. During the period of reflectivity maximum, the tornado traveled almost exclusively through populated areas in Maryland, including College Park, Beltsville, and Laurel, with significant amount of debris being generated. Maximum reflectivities in the knob were 60 dBZ, rivaling reflectivities observed in the parent supercell.

The knob feature has long been associated with tornado-producing hook echoes, showing tornadic velocity signatures (Lemon et al. 1982, Burgess et al. 2002). Burgess et al. (2002) postulated that the localized reflectivity maximum in the knob of the hook is the primary result of large amount of debris being lofted by the tornado. The existence of high reflectivities in the knob of the hook echo, in this case, signifies the presence of a tornado that was inflicting significant amount of damage to structures.

Similar to KLWX WSR-88D radar, KBWI TDWR reflectivity data also displays a well-defined hook echo associated with the supercell throughout most of its lifetime (Fig. 9). The higher resolution of the TDWR system afforded a more detailed display of reflectivity signatures associated with the tornado (i.e., hook echo, high reflectivity knob). Especially apparent was the "wrapping up" of the hook echo as the supercell was in transition from the mature to the collapse stage and tracked from District of Columbia into Maryland between 2113 and 2118 UTC. Also evident was the rapid development and intensification of the high reflectivity "knob" between 2118 and 2128 UTC. During this time, the tornado touched down and rapidly intensified as it tracked from College Park to Beltsville. The well-defined appearance of the hook echo and "knob" was maintained for the remainder of the tornado's lifetime, until approximately 2147 UTC.

At the end of the lifetime of the tornado, TDWR reflectivity imagery observed the evolution of the parent supercell to a bow echo (Figs. 9g and 9h). Fujita (1978) defined the bow echo as a "bow or crescent"-shaped radar echo with a tight reflectivity gradient on the convex (leading) edge, the evolution and horizontal structure of which is consistent with outflow-dominated systems. Bow echoes are typically associated with downbursts, defined by Fujita as strong downdrafts that result in an outburst of damaging winds on or near the earth's surface. By 2152 UTC, reflectivity imagery indicated that the hook echo, now in the vicinity of Columbia, was becoming less defined. At the same time, a weak echo channel was developing on the western edge of the parent supercell, signifying the development of a downburst. By 2157 UTC, the hook echo had dissipated. In addition, the supercell had evolved into a linear pattern that displayed the following radar echo characteristics typical of a distinctive bow echo (Przybylinski and Gery 1983): (a) concave-shaped echo configuration; (b) strong low-level reflectivity gradient along the leading edge of the concave-shaped echo; and (c) a weak echo channel. A distinctive bow echo at this time indicated the presence of damaging downburst winds, the strongest occurring in the vicinity of the weak echo channel near the bow center. Klimowski et al. (2000) identified 21 observed cases of supercell to bow echo evolution and noted that the bows were associated with both severe winds and very large hail.

In the analysis of the 3 May 1999 Oklahoma City tornado, Burgess et al. (2002) defined the maximum velocity difference in the tornado vicinity, across a horizontal distance of less than 1.85 km, and identified a small-scale couplet referred to as a tornado cyclone signature (TCS). They found that the TCS, at low elevation angles, for which the radar beam was less than 1 km AGL, corresponded well to the tornado location, and that the trend in its magnitude was similar to that of the tornado F-scale. The term of tornadic cyclone has been applied to circulations larger than the tornado (Burgess et al. 2002). They also noted that in most observations of tornadoes, such as those collected by the WSR-88D radar, the vortex core is narrower than the effective radar beamwidth. The signature of the tornado in the radar data thus depends greatly on the characteristics of the flow surrounding the core and on the positions of the radar beam relative to the vortex center.

Burgess et al. (2002) defined the quantity Delta-V as the TCS velocity difference. This strong correspondence was demonstrated well in the present case. As displayed in Fig.10, by 2116 UTC, the Delta-V had rapidly exceeded 40 m s$^{-1}$ as the tornado was developing southwest of College Park and then continued to increase to near 45 m s$^{-1}$ by 2121 UTC as the tornado reached F3 intensity. Peak Delta-V of greater than 50 m s$^{-1}$ was observed at 2126 UTC (Figure 10d) as the tornado was tracking rapidly northeastward toward Beltsville. After 2130 UTC, Delta-V decreased to 40 m s$^{-1}$ as the intensity of the tornado weakened to F2. By 2146 UTC (Fig. 10f), the tornado, moving northeast toward Columbia, was near the end of its lifetime and its intensity had decreased to F1. Consequently, Delta-V had decreased significantly to 35 m s$^{-1}$. It was also observed that as the tornado intensified from F1 to F3, the TCS moved from a location on the left front quadrant of the hook to the center of the knob.

The time-height section of TCS Delta-V (Fig.11) reveals significant values before tornado development (> 35 m s$^{-1}$) with higher values aloft (2-4 km AGL) and lower ones near the surface. At the time of tornado development (i.e., 2116 UTC), significant values of Delta-V had formed through a deep column with maximum values (> 40 m s$^{-1}$) observed much closer to the surface (~ 1 km AGL).  By the time of peak tornado intensity (i.e., 2121 UTC), maximum Delta-V had increased to near 50 m s$^{-1}$ and had lowered to near the surface.  After 2136 UTC, weakening of the tornado was apparent as maximum values of Delta-V and the height of the column through which large values had formed had decreased significantly.

      The supercell storm that produced the College Park tornado was first seen at 1915 UTC (Fig. 2a).  By 2015 UTC (not shown) there was an overshooting top associated with the supercell, indicative of intense, deep convection.  At this time, the storm was splitting over Stafford County, Virginia, and was well within the right entrance region of the jet core.  The right-flank supercell continued to track northeastward, steered by strong southwesterly flow aloft and the right entrance region of the jet maximum, and eventually produced the F3 College Park tornado.  At 2015 UTC, the overshooting top was much more apparent as it was casting a shadow on the surrounding cloud mass.  The development of the overshooting top was signifying the further intensification of convection associated with the supercell and the possibility that the storm was becoming severe, producing large hail and damaging winds.  At 2115 UTC (Fig. 2b), the supercell had attained its largest areal extent with the most distinctive overshooting top of its lifetime.  This was an indication that the storm was near its peak intensity with very strong updrafts.  Note the elongated shape of the supercell in the southwest to northeast direction (Fig. 2b) that is indicative of the strong shearing aloft associated with the jet stream.  Based on analysis of the visible imagery, two major conclusions can be stated.  First, intense, deep convection and supercellular activity can be related to the enhanced upward motion in the jet core.  Second, the overshooting top was an important

characteristic of deep and intense convective morphology. We may conclude that the storm updraft rotation was correlated closely to the existence of an overshooting top.

The reflectivity cross section also displays signatures that are highly indicative of tornadic activity (Fig. 12). A distinctive bounded weak echo region (BWER) was present during the lifetime of the supercell. A complete description of the three stages in the evolution of a tornadic supercell are featured in Rotunno (1986). At 2111 UTC (Fig. 12a), the mature supercell displayed the following characteristics: the echo top was at its maximum height (~10 km) with a well-defined BWER, suggesting the existence of a strong updraft and intensification of the mesocyclone as it was building down to the lower levels. Weisman and Klemp (1986) noted that the storm rotation originates through the tilting of horizontal vorticity inherent in the vertically sheared flow. At this time, a hook echo was developing on the right rear flank as indicated in the 2111 UTC reflectivity image (see Fig. 8a). By 2121 UTC (Fig. 12b), the BWER ceiling had lowered as it began to fill, indicating that the supercell was in a collapsing phase (Rotunno 1986). The 2121 UTC reflectivity (Fig. 8c) reveals that the hook echo was "wrapping up" south and east of the parent cell, signifying high probability of tornadic development. The development of the hook echo and BWER collapse demonstrated the intensification of the supercell downdraft and a surge of the gust front as it became highly contorted in the vicinity of the main updraft. It was also apparent that the location of the tornadic cyclone, to be described in the next section, had moved toward the center of the mesocyclone.

# Chapter 4. Model Prediction

In this chapter, the MM5 forecasts, initialized at 1200 UTC 24 September 2001, are first verified against the available observations and then the model results could be used to understand some non-observable phenomena.

**1. Model verification**

In general, the model captures many observed features in satellite and radar imagery, and surface and upper-air observations. For example, the MM5 reproduced a midlevel, negatively-tilted short-wave trough, most apparent at 700 mb (Fig. 3), a jet stream, most pronounced at 500 mb (not shown), and the presence of a warm and moist tongue of air associated with strong southerly flow over central Maryland and Virginia during the mid-afternoon. Low-level southerly flow can be inferred from the strong north-to-south oriented pressure gradient ahead of an eastward progressing cold front. The cumulus lines apparent over the eastern shore were indicative of a warm, moist and unstable south to southeasterly flow originating from the western Atlantic Ocean and the Gulf Stream (Fig. 2a).

It should be reiterated that the updraft of the supercell storm that would eventually produce the College Park tornado was predominantly "fed" by a warm, moist and positively buoyant air mass from the right front quadrant of the storm. It should be noted again that the low-level southeasterly flow overlain by strong southwesterly flow aloft indicated the presence of strong directional wind shear that was to be a major contributor to the development of storm rotation and the favorability for splitting supercell storms. Also inferred from the visible imagery are veering winds with height, which indicated warm air advection and the resultant large-scale ascent over central Maryland. In addition, Fig. 7 shows a gradual increase in CAPE between 2000 and 2100 UTC while the vertical wind shear increases, especially in the lower levels. They were verifiable by the radiosonde observation (RAOB) at 2100 UTC from nearby Dulles Airport, Virginia (KIAD).

A comparison of the model-predicted and observed radar reflectivity at KBWI reveals a similar feature of a convective line with a bow-echo signature over central Maryland (cf. Figs. 9 and 13). This bow-echo signature develops within an hour after the tornado touch down in College Park. The MM5 also reproduces the conditions that were favorable for deep convection in the Washington, DC area during the late afternoon of September 24. A cross section of radar reflectivity, shown in Fig. 12a, shows the development of a deep rotational convective storm that produced the College Park tornado. The west-east cross section through the greater Washington, DC area reveals intense moisture convergence in the boundary layer and upward motion above ahead of a density current. Accordingly, the radar reflectivity cross section displays deep, strong updrafts as indicated by the presence of a bounded weak echo region (BWER). Thus, the effectiveness of the MM5 during the College Park tornado event was underscored by the verification of many of the identified features by satellite, radar and radiosonde observational data.

 2. **Diagnosis of non-observed features**

A major feature identified in 1.33 km resolution, MM5 output was the development and propagation of a solitary wave associated with a density current. Lin and Goff (1988) identified a mesoscale solitary wave in the atmosphere. Their study, based on an event that occurred on 6 March 1969, revealed that solitary waves are propagating features that have the potential to trigger convective storms. Karyampudi et al. (1995) noted that a density current, associated with a Pacific cold front and downslope windstorm, initiated a bore when it descended the Rocky Mountains and impinged on a low-level inversion. The authors of this study highlighted the role of the bore as an effective mechanism for releasing convective instability through parcel lifting.

In this case, the density current-produced solitary wave developed over the Appalachian Mountains during the afternoon of 24 September, and then progressed rapidly eastward into the Maryland and Virginia Piedmont region. Vertical cross sections of potential temperature ($\theta$), given in Figure 14, show the presence of a solitary wave over the Shenandoah Valley at 2000 UTC. Meanwhile, strong downslope (westerly) winds, behind the cold front, had resulted in the formation of a stationary mountain wave just east of the Appalachian Mountains. Since no low-level inversion was present east of the Appalachian Mountains, the density current retained its identity for its entire period of motion. By 2100 UTC, the solitary wave had moved rapidly eastward and was well east of the Blue Ridge Mountains (see Fig. 14b). At 2200 UTC, the solitary wave was located just to the west of the Washington, DC metropolitan area (see Fig. 14c). This solitary wave is a feature of major interest in this study due to its possible role in triggering deep convection during the afternoon of 24 September. The solitary wave most likely released potential instability through rapid vertical lifting. The strong potential instability and vertical wind shear in place over central Maryland and Virginia resulted in the favorability of this deep convection to evolve into supercellular activity.

# Chapter 5.  Summary and Conclusions

The 24 September 2001 College Park, Maryland, tornado was remarkable because of its long track that passed within close range of two Doppler radars and because of its associated features that were identifiable on a number of spatial and temporal scales in satellite, radar and radiosonde observational data as well as numerical model predictions. The tornado featured many similarities to previous significant tornado events that resulted in widespread damage in urban areas, e.g., the 3 May 1999 Oklahoma City tornado.

GOES imagery on 24 September revealed many critical elements of the tornadic event; (a) an upper-level negative-tilt shortwave trough over the Ohio Valley; (b) a wide belt of deep, moist convection ahead of the trough over the eastern United States; (c) widespread cirrus over the mid-Atlantic states associated with the jet stream and transverse banding in the cirrus indicating the presence of a wind speed maximum; (d) breaks in the cirrus over central Maryland revealing the presence of widespread cumulus congestus; (e) cumulus lines indicative of low-level southerly flow; (f) overshooting tops associated with the supercell, indicative of intense, deep convection, and the subsequent collapse of the storm top; and (g) the elongated shape of the supercell due to shearing. Strong CVA, low-level WAA, and vertical wind shear could all be inferred from the satellite imagery.  Also apparent in satellite imagery was strong instability and positive buoyancy that was present over central Maryland to "feed" deep convection.  Satellite imagery animation was utilized to observe the supercell on the storm scale as well as to monitor its evolution as it tracked into a convectively unstable environment.  A striking observation in the imagery was the intensification of the supercell as it moved into a progressively unstable environment and then the collapse of the storm top at the approximate times that the supercell produced a tornado.

In a similar manner to research pertaining to the 3 May 1999 Oklahoma City tornado, operational WSR-88D and TDWR signatures in reflectivity and velocity were shown to have utility in detecting and monitoring the College Park tornado. The radar data revealed (a) a well-defined hook echo; (b) a high reflectivity knob within the hook echo; (c) the evolution of the parent storm cell from a supercell structure to a bow echo; (d) a tornado cyclone signature (TCS) in the velocity data; (e) a large TCS velocity difference (> 40 m s$^{-1}$) at the approximate time of tornado touch-down; and (f) collapse of the bounded weak echo region (BWER) at the time of tornado development. In this case, the high-reflectivity knob of the hook echo was used to infer the presence of a damaging tornado (Burgess et al. 2002). Similar to the observations in the Oklahoma City tornado, the 24 September velocity signatures demonstrated a relationship to the strength of the flow surrounding the tornado. The TCS was effective in indicating the tornado location as well as relative tornado strength. Finally, the evolution of the parent supercell to a bow echo signature was inferred as an indicator of the possible development of a downburst at the end of the tornado's lifetime.

The MM5 forecast, initialized at 1200 UTC 24 September 2001, captured many observed key factors on the synoptic scale and mesoscale during the development of severe convection and the College Park tornado. The forecast also identified an important feature, a solitary wave associated with a density current-cold front that demonstrated a role in the triggering of deep convection during the afternoon of 24 September. Thus, in an operational sense, MM5 data was found to be effective in identifying favorable conditions for severe convection, especially the development of tornadic supercell storms, several hours before the College Park tornado event.

In conclusion, the coordinated use of numerical model data, satellite imagery, and radar data at multiple spatial scales should have provided the operational forecaster with the means to accurately predict the outbreak of severe convection as well as the development of the tornado that tracked through College Park during the afternoon of 24 September 2001. Numerical model guidance provided a long-term outlook of the conditions favorable for the development of severe convection several hours prior to the tornado event while regional satellite imagery proved to be effective in indicating the existence of these conditions one to three hours prior to tornado touch-down in College Park. MM5 data and satellite imagery identified a combination of forcing mechanisms that resulted in the development of the supercell that produced the College Park tornado. Finally, radar imagery proved to be an effective tool in monitoring the structural evolution of the supercell as well as the tornado that tracked through College Park.

# REFERENCES

Bikos, D., J. Weaver, and B. Motta, 2002: A satellite perspective of the 3 May 1999 Great Plains tornado outbreak within Oklahoma. *Wea. Forecasting*, **17**, 635-646.

Bluestein, H.B., 1993: Synoptic-Dynamic Meteorology in Midlatitudes, Volume II, Observations and Theory of Weather Systems. Oxford University Press, Inc., New York, 594pp.

Burgess, D.W., M.A. Magsig, J. Wurman, D. C. Dowell and Y. Richardson, 2002: Radar observations of the 3 May 1999 Oklahoma City tornado. *Wea. Forecasting*, **17**, 456-471.

Chaston, P. R., 1995: Weather Maps. Chaston Scientific, Inc., Kearney, MO, 167pp.

Conway, E.D., 1997: An Introduction to Satellite Image Interpretation. Johns Hopkins University Press, Baltimore, 242pp.

Foster, M.P., A.R. Moller, J.K. Jordan and K. C. Crawford, 2000: Evolution of the surface meteorological fields on May 3, 1999. Preprints, *20th Conf. On Severe Local Storms*, Orlando, FL, Amer. Meteor. Soc., 13-16.

Fujita, T.T., 1978: Manual of downburst identification for project NIMROD. SMRP Research Paper 156, University of Chicago, 104 pp.

Hess, S.L., 1959: Introduction to Theoretical Meteorology. Krieger Publishing Company, Malabar, FL, 362 pp.

Karyampudi, V. M., S.E. Koch, C. Chen, and J.W. Rottman, 1995: The influence of the Rocky Mountains on the 13-14 April 1986 severe weather outbreak. Part II: Evolution of a prefrontal bore and its role in triggering a squall line. *Mon. Wea. Rev.*, **123**, 1423-1446.

Klimowski, B.A., R. Przybylinski, G. Schmocker, and M.R. Hjelmfelt, 2000: Observations of the formation and early evolution of bow echoes. Preprints, *20th Conf. On Severe Local Storms*, Orlando, FL, Amer. Meteor. Soc., 44-47.

Lemon, L. R., D. W. Burgess, and L. D. Hennington, 1982: A tornado extending to great heights as revealed by Doppler radar. Preprints, *12th Conf. on Severe Local Storms,* San Antonio, TX, Amer. Meteor. Soc., 430–432.

Lin, Y., and R.C. Goff, 1988: A study of a mesoscale solitary wave in the atmosphere originating near a region of deep convection. *J. Atmos. Sci.*, **45**, 194-205.

McGinley, J., 1986: Nowcasting Mesoscale Phenomena. In Mesoscale Meteorology and Forecasting. P.S. Ray (Ed.), American Meteorological Society, Boston, 657-688.

Przybylinski, R.W., and W.J. Gery, 1983: The reliability of the bow echo as an important severe weather signature. Preprints, *13th Conf. on Severe Local Storms*, Tulsa, OK, Amer. Meteor. Soc., 270-273.

Rinehart, R.E., 1997: Radar for Meteorologists. Rinehart Publications, Grand Forks, ND, 428pp.

Roebber, P.J., D.M. Schultz, and R. Romero, 2002: Synoptic regulation of the 3 May 1999 tornado outbreak. *Wea. Forecasting*, **17**, 399-429.

Rotunno, R., 1986: Tornadoes and Tornadogenesis. In Mesoscale Meteorology and Forecasting. P.S. Ray (Ed.), American Meteorological Society, Boston, 414-436.

Weber, E.M., and S. Wilderotter, 1993: Satellite Interpretation. In Satellite Imagery Interpretation for Forecasters, P.S. Parke (Ed.), National Weather Association, Washington, DC, 2-D-1-8.

Weisman, M.L., and J.B. Klemp, 1982: The dependence of numerically simulated convective storms on vertical wind shear and buoyancy. *Mon. Wea. Rev.*, **110**, 504-520.

Weisman, M.L., and J. B. Klemp, 1984: The structure and classification of numerically simulated convective storms in directionally varying wind shears. *Mon. Wea. Rev.*, **112**, 2479-2498.

Weisman, M.L. and J.B. Klemp, 1986: Characteristics of Isolated Convective Storms. In Mesoscale Meteorology and Forecasting. P.S. Ray (Ed.), American Meteorological Society, Boston, 331-358.

Weiss, S.J., and D.J. Stensrud, 2000: Mesoscale model ensemble forecasts of the 3 May 1999 tornado outbreak. Preprints, *20th Conf. On Severe Local Storms*, Orlando, FL, Amer. Meteor. Soc., 17-20.